\documentclass[prodmode,acmtecs]{acmsmall} 

\acmVolume{0}
\acmNumber{0}
\acmArticle{0}
\acmYear{0000}
\acmMonth{0}

\usepackage[table]{xcolor}
\usepackage{amsmath,amssymb}
\usepackage{threeparttable}
\usepackage{tabulary}
\usepackage{multirow}
\usepackage{subfigure}
\usepackage{mathtools} 
\usepackage{tikz}
\usetikzlibrary{arrows,automata,positioning}
\usepackage[ruled,vlined]{algorithm2e}

\usepackage{xspace}
\usepackage{wrapfig}
\usepackage{caption}
\usepackage{hyperref}
\usepackage{bm}

\newcommand{\cceq}{\mathop{::=}}

\newcommand{\nmodels}{\nvDash}

\begin{document}
\newcommand{\comp}[1]{\textsf{\small #1}\xspace}
\newcommand{\pow}[1]{2^{#1}}
\newcommand{\nats}{\mathbb{N}}
\newcommand{\size}[1]{|#1|}
\newcommand{\set}[1]{\{ #1 \}}
\newcommand{\ap}[0]{\mathrm{AP}}
\newcommand{\notleftright}{\mathrel{\ooalign{$\leftrightarrow$\cr\hidewidth$/$\hidewidth}}}
\newcommand{\true}{\mathit{true}}
\newcommand{\false}{\mathit{false}}
\newcommand{\F}{{\mathbf{F\,}}}
\newcommand{\G}{{\mathbf{G\,}}}
\newcommand{\U}{{\mathbf{\,U\,}}}
\newcommand{\X}{{\mathbf{X\,}}}
\newcommand{\R}{{\mathbf{\,R\,}}}
\newcommand{\Waitfor}{{\mathbf{\,W\,}}}
\newcommand\equ{\leftrightarrow}
\newcommand{\ldot}{\mathpunct{.}}
\newcommand{\nat}{\mathbb{N}}

\newcommand{\ltl}{{LTL}\xspace}
\newcommand{\hyltl}{{HyperLTL}\xspace}
\newcommand{\hyctlstar}{{HyperCTL$^*$}\xspace}
\newcommand{\fol}{{FO}$[<]$\xspace}
\newcommand{\fole}{{FO}$[<,\,E]$\xspace}
\newcommand{\hyfol}{{HyperFO}\xspace}
\newcommand{\kltl}{{KLTL}\xspace}
\newcommand{\fo}{FO[$<$]\@\xspace}
\newcommand{\foe}{\mbox{FO[$<,E$]}\@\xspace}
\newcommand{\seins}{S1S[$E$]\@\xspace}
\newcommand{\msoe}{MSO[$E$]\@\xspace}
\newcommand{\mple}{MPL[$E$]\@\xspace}

\newcommand{\K}{{\mathbf{\,K\,}}}
\newcommand{\suffix}[2]{#1[#2,\infty]}
\newcommand{\var}{\mathcal{V}}

\newcommand{\phiswap}{\phi_{\mathrm{swp}}}
\newcommand{\phiproper}{\phi_{\mathrm{prp}}}
\newcommand{\bal}{\mathrm{bal}}

\newcommand{\pathassign}{\Pi}
\newcommand{\Tra}{\ensuremath{\mathit{Tr}}}
\newcommand{\Paths}{\ensuremath{\mathit{Paths}}}
\newcommand{\donotshow}[1]{}

\newcommand{\Veins}{\mathcal{V}_1}
\newcommand{\Vzwei}{\mathcal{V}_2}


\title{Logics and Algorithms for Hyperproperties}
\author{Bernd Finkbeiner
\affil{CISPA Helmholtz Center for Information Security}
}
\begin{abstract}
  System requirements related to concepts like information flow,
  knowledge, and robustness cannot be judged in terms of individual
  system executions, but rather require an analysis of the
  relationship between multiple executions.  Such requirements belong
  to the class of hyperproperties, which generalize classic trace
  properties to properties of sets of traces.  During the past decade,
  a range of new specification logics has been introduced with the goal
  of providing a unified theory for reasoning about hyperproperties.
  This paper gives an overview on the current landscape of logics for the
  specification of hyperproperties and on algorithms for satisfiability
  checking, model checking, monitoring, and synthesis.
\end{abstract}

\maketitle

\section{Introduction}

In 2008, Clarkson and Schneider coined the term \emph{hyperproperties}
for the general class of properties that relate
multiple executions of a computer system~\cite{4556678}. In many branches of computer science, however,
the study of hyperproperties began long before that.  The literature on \emph{information flow security} contains
many variations of the \emph{noninterference} property, which requires
that for all computations and all sequences of actions of a
high-security agent $A$, the resulting observations made by a
low-security observer $B$ are identical to $B$'s observations that
would result without $A$'s
actions~\cite{GoguenM/1982/Noninterference}.  Noninterference has been
adapted to a variety of scenarios, attackers, and system models
(cf.~\cite{sabelfeld07JCS,mantel07ESOP}).  In the design of
\emph{distributed systems}, information flow and the resulting
\emph{knowledge} of agents plays a fundamental role~\cite{FHMV,HM}.
Another example of a hyperproperty that has been studied in many
disciplines is the engineering notion of
\emph{robustness}~\cite{Dullerud}, i.e., the guarantee that a system
will not deviate strongly from its expected trajectory when
disturbances enter the system. Variations of robustness have been
applied to a wide range of software systems, from embedded
controllers~\cite{10.1007/978-3-319-47169-3_46} to sorting
algorithms~\cite{DBLP:journals/cacm/ChaudhuriGL12}.

Despite the strong interest in hyperproperties across many different research areas, early proposals for specification languages were limited to hyperproperties from particular domains. A branch of hyperproperties that has seen active logic development are properties related to knowledge, such as the \emph{temporal logic of knowledge}~\cite{FHMV}. Reasoning about \emph{combinations of programs} is possible in \emph{relational logics}, such as \emph{relational Hoare Logic} (RHL), originally introduced by Benton~\cite{Benton04}. Generally, relational logics are restricted to properties that express a condition over a given set of programs or a $k$-fold self-composition of some program for some fixed $k$; there are, however, extensions that are directed at specific properties such as differential privacy~\cite{BartheKOB13} and sensitivity~\cite{BartheEGHS18}.

A significant step towards a general logic for hyperproperties was the
introduction of \emph{HyperLTL}~\cite{DBLP:conf/post/ClarksonFKMRS14}.
HyperLTL adds universal and existential quantification over traces to
standard propositional linear-time temporal logic (LTL). HyperLTL
can complex information-flow properties like
generalized noninterference, declassification, and quantitative
noninterference. Unlike specialized techniques for particular
hyperproperties, HyperLTL provides a unifying logic for expressing and
verifying a wide range of policies.  A key benefit of such a  logic is that it can serve as a common input
language and semantic framework for a range of algorithmic tools. In the paper,
we provide an overview on algorithms for satisfiability
checking, model checking, monitoring, and synthesis with HyperLTL as
the common specification language.

While HyperLTL has found many applications, it is perhaps not
surprising that there are also numerous hyperproperties 
that cannot be expressed in HyperLTL. An early observation was that
the expressiveness of HyperLTL is incomparable to epistemic temporal
logic~\cite{BozzelliMP15}. To address this lack of
expressiveness in HyperLTL, a promising idea is to replace LTL  with a
more powerful logic as the underlying trace logic of HyperLTL.

In the realm of trace properties, LTL is the starting point of a
well-known hierarchy of logics, where LTL is equivalent, by Kamp's
theorem~\cite{Kamp68}, to monadic first-order logic of
order \fo. Extending LTL with quantification over propositions leads
to the strictly more expressive QPTL~\cite{QPTL}, which
can express all $\omega$-regular properties and is also expressively
equivalent to monadic second-order logic of one successor
(S1S)~\cite{QPTL-S1S}. Extending LTL to branching time leads to the
temporal logic CTL$^*$, which has been shown to be expressively equivalent
to monadic path logic (MPL)~\cite{abrahamson1980}. Adding quantification over propositions leads
to the strictly more expressive quantified computation tree logic (QCTL$^*$)~\cite{QCTLStar}, which is expressively equivalent to monadic second-order logic (MSO)~\cite{laroussinie2014}.

\begin{figure}
	\centering
	\begin{tikzpicture}[lblell/.style={ellipse,fill=white, fill opacity=0.3, text opacity=1},
	lblrect/.style={rectangle,fill=black, fill opacity=0.3, text opacity=1},
	scale=1]
	{\draw[black,line width =0.5mm] (0,0) circle (2.5cm);}
	{\draw[black,  fill=gray!60, fill opacity = 0.3,line width =0.5mm] (0,-0.5) circle (2 cm);}
	{\draw[black, line width =0.5mm] (0,-1) circle (1.5 cm);}
	{\draw[black,line width =0.5mm] (0,-1.5) circle (1 cm);}
	{\node (hyperltl) at (0,-1.5) {HyperLTL};}
	{\node (hyperltl) at (0,-0.1) {FO[$<$,$E$]};}
	{\node (hyperltl) at (0,0.9) {HyperQPTL};}
	{\node (hyperltl) at (0,2) {S1S[$E$]};}
	\end{tikzpicture}\qquad
\begin{tikzpicture}[lblell/.style={ellipse,fill=white, fill opacity=0.3, text opacity=1},
	lblrect/.style={rectangle,fill=black, fill opacity=0.3, text opacity=1},
	scale=1]
	{\draw[black,line width =0.5mm] (0,0) circle (2.5cm);}
	{\draw[black,  fill=gray!60, fill opacity = 0.3,line width =0.5mm] (0,-0.75) circle (1.75 cm);}
	{\draw[black,line width =0.5mm] (0,-1.5) circle (1 cm);}
	{\node (hyperltl) at (0,-1.5) {HyperCTL$^*$};}
	{\node (hyperltl) at (0,0.15) {MPL[$E$]};}
	{\node (hyperltl) at (0,1.65) {\parbox[c]{2cm}{\centering HyperQCTL$^*$\\ = MSO[$E$]}};}
	\end{tikzpicture}
        
	\caption{The hierarchy of hyperlogics~\protect\cite{hierarchy_hyperlogics}. Linear-time logics are shown on the left, branching-time logics on the right. The shaded areas indicate logics with decidable model-checking problem.}
	\label{fig:hier}
\end{figure}
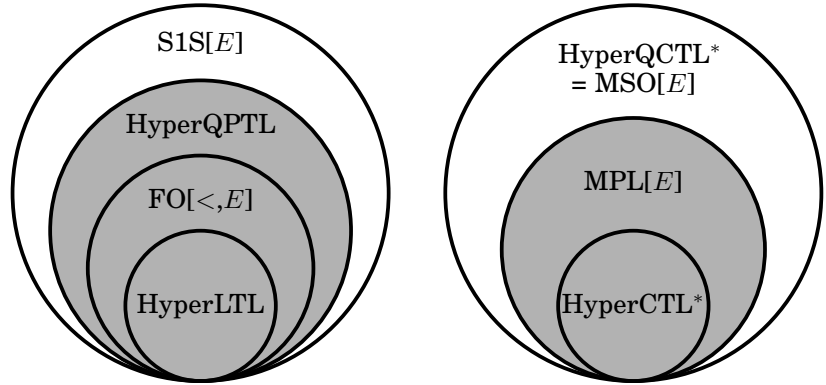

Using the extension of LTL to HyperLTL as a blueprint, we extend the temporal logics QPTL, CTL$^*$, and QCTL$^*$ with quantification over traces or paths to obtain the corresponding hyperlogics. A different method for the construction of hyperlogics has been introduced for first-order and second-order logics. Here, the extension consists of adding the \emph{equal-level predicate}~$E$ (cf.~\cite{Thomas09,Martin}), which relates the same time points on different traces.
Figure~\ref{fig:hier} shows the resulting hierarchy of hyperlogics (introduced in~\cite{hierarchy_hyperlogics}).
For linear time, the most striking difference to the hierarchy of the standard logics is that, when lifted to hyperlogics, \fo is no longer equivalent to LTL, and S1S is no longer equivalent to QPTL: \seins is strictly more expressive than HyperQPTL; HyperQPTL, in turn, is strictly more expressive than \foe, and \foe is strictly more expressive than HyperLTL.
For branching time, we have that HyperQCTL$^*$ is still expressively equivalent to MSO$[E]$. 
However, MPL, which is equivalent to CTL$^*$ in the standard hierarchy, falls strictly between HyperCTL$^*$ and HyperQCTL$^*$ when equipped with the equal-level predicate (\mple).

The logics in the hierarchy significantly extend the expressiveness of HyperLTL. Already the extension to QPTL suffices, for example, to subsume epistemic temporal logic~\cite{markus}. At the same time, a large part of the hierarchy is still supported by algorithmic analysis. The shaded areas of Fig.~\ref{fig:hier} indicate logics where the model-checking problem is decidable.

The focus of the next sections is on HyperLTL. We introduce the syntax and semantics of the logic in Section~\ref{sec:hltl} and discuss the satisfiability, model checking, monitoring, and synthesis problems in Sections~\ref{sec:sat}--\ref{sec:synthesis}. We return to the hierarchy of hyperlogics in Section~\ref{sec:hier}, where we also briefly discuss further extensions to probabilistic, infinite-state, and asynchronous hyperproperties.

\section{HyperLTL}
\label{sec:hltl}
        
HyperLTL~\cite{DBLP:conf/post/ClarksonFKMRS14} is an extension of linear-time temporal logic (LTL)~\cite{Pnueli/1977/TheTemporalLogicOfPrograms}. In the following, we briefly define the syntax and semantics of LTL and then describe the extension to HyperLTL.
Let $\ap$ be a finite set of atomic propositions. A trace over $\ap$ is a map $t \colon \nats \rightarrow \pow{\ap}$, denoted by $t(0)t(1)t(2) \cdots$. Let $(\pow{\ap})^\omega$ denote the set of all traces over $\ap$.

\subsection{LTL}
The formulas of LTL
are generated by the following grammar: \vspace{-1mm}
$$
\varphi~~::=~~ a ~~|~~ \neg\varphi ~~|~~ \varphi\wedge\varphi ~~|~~ \X \varphi ~~|~~ \varphi \U \varphi \vspace{-.5mm}
$$
where $a\in\ap$ is an \emph{atomic proposition}, the Boolean connectives $\neg$ and $\wedge$ have the usual meaning, $\X$ is the temporal \emph{next} operator, and $\U$ is the temporal \emph{until} operator. 
We also consider the usual derived Boolean connectives, such as $\vee$, $\rightarrow$, and $\equ$, and the derived temporal operators \emph{eventually} $\F\varphi\equiv \true\U\varphi$, \emph{globally} $\G\varphi\equiv\neg\F\neg\varphi$, \emph{weak until} $\varphi \Waitfor \psi \equiv (\varphi \U \psi) \vee \G \varphi$, and \emph{release} $\varphi \R \psi \equiv \neg (\neg \varphi \U \neg \psi)$.
The satisfaction of an LTL formula $\varphi$ over a trace $t$ at a position $i\in \mathbb N$, denoted by $t,i \models \varphi$, is defined as follows:
\[\def\arraystretch{1.1}
\begin{array}{l@{\hspace{1em}}c@{\hspace{1em}}l}
  t,i\models a & \text{iff} &  a\in t(i),\\
  t,i \models \neg \varphi & \text{iff}&  t,i \not\models \varphi,\\
  t,i\models \varphi_1 \wedge \varphi_2 & \text{iff} &  t,i \models \varphi_1 \text{ and } t,i \models \varphi_2,\\
  t,i\models\X\varphi & \text{iff} &  t,i+1\models\varphi,\\
  t,i\models \varphi_1\!\U\!\varphi_2 & \text{iff} & \exists k \geq i :~t,k \models\varphi_2~\wedge~\forall i \leq j < k:~t,j \models\varphi_1.
\end{array}
\]
We say that a trace $t$ satisfies a sentence~$\phi$, denoted by $t \models \phi$, if $t,0 \models \phi$.
For example, the LTL formula $\G(a\rightarrow\F b)$ specifies that every position in which $a$ is true must eventually be followed by a position where $b$ is true.


\subsection{HyperLTL}
The formulas of \hyltl are generated by the grammar 
\begin{align*}
\phi & {} \cceq {}  \exists \pi.\ \phi ~\mid~ \forall \pi.\ \phi ~\mid~ \psi \\
\psi & {}  \cceq {}  a_\pi ~\mid~ \neg \psi ~\mid~ \psi \wedge \psi ~\mid~ \X \psi ~\mid~ \psi \U \psi 
\end{align*}
where $a$ is an atomic proposition from a set $\ap$ and $\pi$ is a trace variable from a set $\var$. Further Boolean connectives and the temporal operators $\F$, $\G$, $\Waitfor$, and $\R$ are derived as for LTL. 

The semantics of \hyltl is defined with respect to a trace assignment, a partial mapping~$\Pi \colon \var \rightarrow (\pow{\ap})^\omega$. The assignment with empty domain is denoted by $\Pi_\emptyset$. Given a trace assignment~$\Pi$, a trace variable~$\pi$, and a trace~$t$, we denote by $\Pi[\pi \rightarrow t]$ the assignment that coincides with $\Pi$ everywhere but at $\pi$, which is mapped to $t$.
%
The satisfaction of a HyperLTL formula $\phi$ over a trace assignment $\Pi$ and a set of traces $T$ at a position $i \in \mathbb N$, denoted by $T,\Pi,i \models \phi$, is defined as follows:

\[\def\arraystretch{1.1}
\begin{array}{l@{\hspace{1em}}c@{\hspace{1em}}l}
  T, \Pi,i \models a_\pi & \text{iff} & a \in \Pi(\pi)(i),\\
  T, \Pi,i \models \neg \psi & \text{iff} & T, \Pi,i \nmodels \psi,\\
  T, \Pi,i \models \psi_1 \wedge \psi_2 & \text{iff} & T, \Pi,i \models \psi_1 \text{ and } T, \Pi,i \models \psi_2,\\
  T, \Pi,i \models \X \psi & \mbox{iff} & T,\Pi,i+1 \models \psi,\\
  T, \Pi,i \models \psi_1 \U \psi_2 & \text{iff} &  \exists k \ge i.\ T,\Pi, k \models \psi_2\\
  & & \wedge \forall i \le j < k.\ T,\Pi,j \models \psi_1,\\
  T, \Pi,i \models \exists \pi.\ \phi & \text{iff} & \exists t \in T.\ T,\Pi[\pi \rightarrow t],i \models \phi,\\
  T, \Pi,i \models \forall \pi.\ \phi & \text{iff} & \forall t \in T.\ T,\Pi[\pi \rightarrow t],i \models \phi.
  \end{array}
\]
We say that a set $T$ of traces satisfies a sentence~$\phi$, denoted by $T \models \phi$, if $T, \Pi_\emptyset,0 \models \phi$.

	A \emph{Kripke structure} is a tuple $\mathcal K=(S,s_0,\delta,\ap,L)$ consisting of a set of states $S$, an initial state $s_0$, a transition function $\delta:S\to 2^{S}$, a set of \emph{atomic propositions} $\ap$, and a \emph{labeling function}  $L:S\to 2^{\ap}$ that assigns to each state a set of atomic propositions that are true in the state.
	We require that each state $s$ has a successor, that is $\delta(s)\neq\emptyset$, to ensure that every execution of a Kripke structure can always be continued to infinity.
        In a \emph{finite} Kripke structure, $S$ is a finite set. 

A \emph{path} of a Kripke structure is an infinite sequence $s_0s_1\ldots\in S^\omega$ such that $s_0$ is the initial state of $\mathcal K$ and $s_{i+1}\in\delta(s_i)$ for all $i\in \mathbb{N}$. 
By $\Paths(\mathcal K,s)$, we denote the set of all paths of $\mathcal K$ starting in state $s\in S$.
  A \emph{trace} of a path $\sigma=s_0s_1\ldots$ is a sequence of labels $l_0l_1\ldots$ with $l_i=L(s_i)$ for all $i\in\mathbb{N}$. $\Tra(K,s)$ is the set of all traces of paths of a Kripke structure $\mathcal K$ starting in state~$s$. 
  A Kripke structure $\mathcal K$ with initial state $s_0$ satisfies an LTL formula $\varphi$, denoted by $\mathcal K \models \varphi$ iff for all traces $\pi \in \Tra(\mathcal K,s_0)$, it holds that $\pi \models \varphi$. Likewise, the Kripke structure satisfies a HyperLTL formula $\phi$, also denoted by $\mathcal K \models \phi$, iff $\Tra(\mathcal K,s_0) \models \phi$.

\subsection{Examples}

We illustrate the specification of hyperproperties with HyperLTL using some prominent examples from information-flow security.

\begin{itemize}
  \item A system satisfies \emph{observational determinism}~\cite{Zdancewic+Myers/03/ObservationalDeterminism} if every pair of traces with the same low-security input remains indistinguishable for an observer; i.e., the program appears to be deterministic to low-security users.
For a system with low-security input $l$ and output $o$, 
observational determinism can be expressed in HyperLTL as follows:
\[
\forall \pi.\forall\pi'.\ (l_{\pi} \leftrightarrow l_{\pi'}) ~\rightarrow~ \G (o_{\pi} \leftrightarrow o_{\pi'})
\]

\item \emph{Noninference}~\cite{McLean:1994:GeneralTheory} specifies that the behavior observable by an observer must not change when all high-security inputs are replaced by a dummy input. In the following HyperLTL formula, we express noninference for a high-security input $h$, which we set to \emph{false} as dummy input, low-security input $l$, and output $o$: 
\[
\forall\pi.\exists\pi'. \ \G\, (\neg h_{\pi'} \wedge (l_{\pi}\leftrightarrow l_{\pi'}) \wedge (o_\pi\leftrightarrow o_{\pi'}))
\]

\item \emph{Generalized noninterference}~\cite{McCullough:1987:GNI} is similar to observational determinism in that it requires that the output ($o$) should not be influenced by high-security input ($h$). Unlike observational determinism, it allows, however, for nondeterministic behavior in the low-security input ($l$):
\[
\forall\pi.\forall\pi'.\exists\pi''.\ \G\, ( (h_{\pi}\leftrightarrow h_{\pi''}) \wedge (l_{\pi'}\leftrightarrow l_{\pi''}) \wedge (o_{\pi'} \leftrightarrow o_{\pi''}))
\]
The existentially quantified trace $\pi''$ combines the
high-security input of the universally quantified trace $\pi$ with the low-security input and output of the universally quantified trace $\pi'$. 
\end{itemize}

As discussed in the introduction, hyperproperties have many applications beyond security. An illustrative example is the specification of \emph{error resistant codes,} which transmit data over noisy channels (cf.~\cite{DBLP:conf/cav/FinkbeinerRS15}). A typical correctness condition for such a code is that all code
words have a certain minimal Hamming distance. The following HyperLTL formula specifies that all code words produced by an encoder with input $i$ and output $o$ have a Hamming distance of at least $d$:
\[
  \forall\pi.\forall\pi'.\ (\F(i_\pi \!\leftrightarrow\! \neg i_{\pi'})) \rightarrow \neg\text{Ham}(d-1,\pi,\pi')
\]
where the subformula $\text{Ham}(d,\pi,\pi')$ is defined recursively as follows:
\[
\renewcommand{\arraystretch}{1.1}
\begin{array}{ll}
\text{Ham}(-1,\pi,\pi')&\hspace{0pt}= \false\\  
\text{Ham}(d,\pi,\pi')&\hspace{0pt}= (o_\pi\!\leftrightarrow\!o_{\pi'}) \Waitfor ((o_\pi\!\leftrightarrow\!\neg o_{\pi'})\, \wedge\, \X \text{Ham}(d\!-\!1,\pi,\pi')) .
\end{array} 
\]
The subformula $\text{Ham}(d,\pi,\pi')$ expresses that the Hamming distance between the outputs on $\pi$ and $\pi'$ is \emph{less than} $d$: for $d \geq 0$ this means that either $\pi$ and $\pi'$ agree forever on the output $o$, or there is a point in time where they differ and the suffixes from that point onward have a Hamming distance of less than $d-1$. The full specification thus means that on any pair of traces $\pi, \pi'$ where the input eventually differs, the Hamming distance must not be less than $d$.

\donotshow{ 

  \paragraph{Expressiveness.} In addition to the examples already given in the introduction, two interesting hyperproperties expressible in HyperLTL are security policies based on quantitative information-flow and the minimal Hamming distance of error-resistant codes. The following HyperLTL encodings of these examples are taken from \cite{DBLP:conf/post/ClarksonFKMRS14}, where further details and more examples can be found.

\emph{Quantitative information-flow} policies limit the leakage of information to a certain rate. The following HyperLTL formula expresses that there is no tuple of $2^n+1$ low-distinguishable traces (cf.~\cite{Smith/2009/OnTheFoundationsOfQantitativeInformationFlow,Yasuoka+Terauchi/2010/OnBoundingProblemsOfQuantitativeInformationFlow}):
\[
\forall \pi_0.\;\dots\;.~\forall\pi_{2^n} . ~\Big(\bigvee_{i}\pi_i \neq_{L,\mathsf{in}}\pi_0\Big) ~\vee \bigvee_{i\not=j}\pi_i =_{L,\mathsf{out}}\pi_j
\]
}


\section{Satisfiability}
\label{sec:sat}

\def\arraystretch{1.5}
\begin{table}[t]
    \begin{threeparttable}
  \caption{Complexity of \hyltl satisfiability for formulas from the full logic, and for the temporal safety and liveness fragments. A HyperLTL formula is temporal safety (resp. temporal liveness) if its LTL body describes a safety (resp. liveness) property. All results denote completeness.}
  \label{table:satisfiability}
  \centering
  \begin{tabular}{|c|c|c|c|c|c|c|c|}    \cline{1-2}\cline{4-5}\cline{7-8}
\multicolumn{2}{|c|}{\bf Full HyperLTL} && \multicolumn{2}{|c|}{\bf Temporal Safety} && \multicolumn{2}{|c|}{\bf Temporal Liveness} \\ 
    \cline{1-2}\cline{4-5}\cline{7-8}
full logic & $\Sigma^1_1$~\tnote{1} && full fragment & coRE~\tnote{3} && full fragment & $\Sigma^1_1$~\tnote{3} \\
$\exists^* \forall^*$ & EXPSPACE~\tnote{2} &  & $\forall^*\exists^*\, \G(\mathbf{X}^*)$ & coRE~\tnote{3}& & $\forall\exists^*\, \F (\mathbf{X}^*)$ & NP~\tnote{3} \\
     $\exists^*$ & PSPACE~\tnote{2} & & $\forall\exists^*\, \G$ & NEXP~\tnote{3}&& $\forall\exists^*\, \F \wedge \cdots \wedge \F$ & NP~\tnote{3} \\ \cline{7-8}
$\forall^*$ & PSPACE~\tnote{2} & & $\forall^*\exists^*\, \mathbf{X}^*$ & NEXP~\tnote{3} & \multicolumn{3}{c}{}\\
\cline{1-2} \cline{4-5}
  \end{tabular}
\begin{tablenotes}
\item[1] \cite{DBLP:conf/mfcs/FortinKT021}
\item[2] \cite{DBLP:conf/concur/FinkbeinerH16}
\item[3] \cite{DBLP:conf/lics/BeutnerCFHK22}
\end{tablenotes}
\end{threeparttable}

\end{table}

In the HyperLTL satisfiability problem, we decide, for a given HyperLTL formula~$\phi$, whether or not there exists a set $T$ of traces such that $T \models \phi$. Table~\ref{table:satisfiability} gives an overview on the complexity of the satisfiability problem. Satisfiability of HyperLTL formulas without quantifier alternation is PSPACE-complete~\cite{DBLP:conf/concur/FinkbeinerH16}, which is the same complexity as LTL satisfiability~\cite{10.1145/3828.3837}. 
For formulas from the $\exists^*\forall^*$ fragment, the complexity increases to EXPSPACE~\cite{DBLP:conf/concur/FinkbeinerH16}.  Formulas of this fragment can be translated into equisatisfiable (but exponentially larger) formulas with only existential formulas by explicitly enumerating all possible interactions between the universal and existential quantifiers. Satisfiability of hyperproperties with $\forall^*\exists^*$ trace quantifier alternation is undecidable \cite{DBLP:conf/concur/FinkbeinerH16,DBLP:conf/mfcs/FortinKT021}.

\subsection{Satisfiability of the $\forall^*\exists^*$ fragment}

The $\forall^*\exists^*$ fragment contains many interesting hyperproperties, such as noninference and generalized noninterference. Despite the general undecidability, some positive results have been obtained by classifying hyperproperties into temporal safety and temporal liveness~\cite{DBLP:conf/lics/BeutnerCFHK22}. A HyperLTL formula is \emph{temporal safety} (resp. \emph{temporal liveness}) if its LTL body describes a safety (resp. liveness) property.\footnote{Temporal safety and temporal liveness differ from the notions of hypersafety and hyperliveness~\cite{4556678}.} The restriction to temporal safety reduces the complexity from $\Sigma^1_1$ to coRE. While still undecidable, this enables using common first-order techniques such as resolution, tableaux, and related methods~\cite{DBLP:books/el/RobinsonV01}. In contrast to temporal safety properties, the class of temporal liveness formulas is of analytical complexity. As shown in Table~\ref{table:satisfiability}, both temporal safety and temporal liveness contain decidable fragments. A notable difference is that while formulas from the $\forall^*\exists^*\, \G(\mathbf{X}^*)$ fragment are undecidable, formulas from the $\forall\exists^*\, \F (\mathbf{X}^*)$ are decidable~\cite{DBLP:conf/lics/BeutnerCFHK22}.

\subsection{Practical algorithms}

Despite the general undecidability, there is significant practical interest in algorithms that detect satisfiability and unsatisfiability of HyperLTL formulas.

\subsubsection*{Decision procedures} The decidability of the $\exists^*\forall^*$ fragment is particularly useful in practice because it allows us to check implications between alternation-free formulas. The tool EAHyper implements the  decision procedure as a reduction to LTL satisfiability~\cite{EAHyper}.

\subsubsection*{Finding largest models} For formulas in the $\forall \exists^*$ fragment, an incomplete method to detect (un)satisfiability is to search for the largest model that satisfies the formula. For satisfiable $\forall \exists^*$ formulas, this model is unique. The algorithm iteratively eliminates choices for the $\exists^*$ quantifiers that admit no witness trace when chosen for the $\forall$ quantifier~\cite{DBLP:conf/lics/BeutnerCFHK22}. 

\subsubsection*{Finding smallest models} A method for general HyperLTL formulas is to search for finite models of bounded size and then iteratively increase the bound~\cite{DBLP:journals/corr/abs-1903-11138,DBLP:conf/csl/Mascle020}. Such an approach finds smallest models, but cannot determine unsatisfiability.

\section{Model Checking}
\label{section:modelchecking}
\def\arraystretch{2}
\begin{table}[t]
  \begin{threeparttable}
\caption{Complexity of the HyperLTL model-checking problem for tree-shaped, acyclic, and general Kripke structures. The complexity  is given in the size of 
the Kripke structure, where $k$ is the number of quantifier alternations in 
$(\forall^*\exists^*)^*$.  All results denote completeness.}
\label{tab:system}
\begin{tabular}{|c||c||c||c|}
\hline
 & {\bf Trees } & {\bf Acyclic graphs } & {\bf General graphs}\\
\hline\hline
$\forall^+/\exists^+$ &  & NL~\tnote{1} & NL~\tnote{2}\\
\cline{1-1}\cline{3-4}

$\exists^+\forall^+$/$\forall^+\exists^+$ &  L~\tnote{1} & 
NP/coNP~\tnote{1} & PSPACE~\tnote{3}\\
\cline{1-1}\cline{3-3}\cline{4-4}

\multirow{2}{*}{$(\forall^*\exists^*)^*$}  &  & 
$\mathsf{\Pi}_{k}^p$~\tnote{1} & 
\multirow{2}{*}{$(k{-}1)$-EXPSPACE~\tnote{4}}\\
\cline{3-3}

 &  & $\mathsf{\Sigma}_{k}^p$~\tnote{1} & \\
\hline
\end{tabular}

\begin{tablenotes}
\item[1] \cite{DBLP:conf/csfw/BonakdarpourF18}
\item[2] \cite{DBLP:conf/cav/FinkbeinerRS15}
\item[3] \cite{DBLP:conf/post/ClarksonFKMRS14}
\item[4] \cite{markus}
\end{tablenotes}
\end{threeparttable}
\end{table}
In model checking, we decide, for a given finite Kripke structure $\mathcal K$ and a given HyperLTL formula $\phi$, whether or not $\mathcal K \models \phi$. The model-checking problem is decidable. As shown in the right-most column of Table~\ref{tab:system}, the complexity is non-elementary in the number of quantifier alternations; each quantifier alternation causes a further exponential cost. Fortunately, many hyperproperties of interest have only a small number of quantifier alternations. Observational determinism, for example, is a universal formula and, therefore, only requires nondeterministic logarithmic space in the size of the Kripke structure; noninference and generalized noninterference both have a single quantifier alternation and, hence, can be checked in PSPACE.
The table furthermore shows that quantifier alternation is much less costly for Kripke structures with restricted structures, such as acyclic graphs and trees.\footnote{To satisfy the requirement that every state of a Kripke structure has a successor state, we assume implicit selfloops on terminal states in tree-shaped and acyclic Kripke structures~\cite{DBLP:conf/csfw/BonakdarpourF18}.} For acyclic graphs, the complexity of all model-checking problems is in the polynomial hierarchy within PSPACE, for trees even in deterministic logarithmic space. In the following, we first sketch a basic automata-theoretic model-checking procedure, and then discuss more practical approaches.

\subsection{The basic algorithm}

The following basic algorithm (described in more detail in~\cite{DBLP:conf/cav/FinkbeinerRS15}) reduces the model-checking problem to the language emptiness problem of a B\"uchi automaton: the given Kripke structure satisfies the formula if and only if the language of the resulting automaton is empty. 

The construction starts by negating the given formula, so that the resulting formula describes the existence of an error. Let the new formula have the form $Q_n \pi_{n} Q_2 \pi_{n-1}\ldots Q_1 \pi_1 .\ \psi$, where $Q_1, Q_2, \ldots Q_n$ are trace quantifiers in $\{ \exists, \forall \}$, and $\psi$ is a quantifier-free formula over atomic propositions indexed by trace variables $\{\pi_1, \ldots \pi_n\}$. Similar to standard LTL model checking, we convert the LTL formula $\psi$  into an equivalent B\"uchi automaton $\mathcal A_0$ over the alphabet $(2^\ap)^n$. Each letter is a tuple of $n$ sets of atomic propositions, where the $i$th element of the tuple represents the atomic propositions of trace $\pi_i$.

The algorithm then eliminates the quantifiers one at a time, starting with the innermost quantifier. Before the elimination of the $i$th quantifier, the automaton $\mathcal A_{i-1}$ over alphabet $(2^\ap)^{(n-i)}$ has been constructed, and now the first component of the tuple corresponds to $\pi_i$.  We assume that $Q_i$ is an existential quantifier; if $Q_i$ is a universal quantifier, we turn it into an existential quantifier by complementing $\mathcal A_{i-1}$, based on the equivalence $\forall \pi. \phi = \neg \exists \neg \phi$.

We then combine $\mathcal A_{i-1}$ with the Kripke structure $\mathcal K$ to ensure that the first component  is chosen consistently with some path in $\mathcal K$. Since $Q_i$ is an existential quantifier, we eliminate the first component of the tuple by existential projection on the automaton. This results in the next automaton $\mathcal A_i$. After $n$ such steps, all quantifiers have been eliminated and the language of the resulting automaton is over the one-letter alphabet consisting of the empty tuple. The HyperLTL formula is satisfied if and only if the language of automaton $\mathcal A_n$ is empty.

\subsection{Practical algorithms}

Because of the expensive automata constructions, it is difficult to implement the basic algorithm efficiently. In the following, we briefly discuss some more practical approaches.

\subsubsection*{Self-composition}
For the alternation-free fragment of HyperLTL, the model-checking problem can be solved by \emph{self-composition}~\cite{DBLP:journals/mscs/BartheDR11}. If the system is given as a circuit $C$ and the HyperLTL formula consists of an LTL formula $\psi$ and a quantifier prefix consisting of $k$ exclusively universal quantifiers (the case of exclusively existential quantifiers is analogous), then a new circuit can be constructed that consists of $k$ copies of $C$ and a monitor circuit $C_\psi$ that recognizes violations of $\psi$. The HyperLTL model-checking problem then reduces to a standard model-checking problem of a trace property on the newly constructed circuit. This approach has been implemented in the HyperLTL model checker MCHyper~\cite{DBLP:conf/cav/FinkbeinerRS15}. 

\subsubsection*{Strategy-based verification} 
HyperLTL properties with a $\forall^*\exists^*$ quantifier alternation can be verified by finding a winning strategy for the ``existential'' player in a game against a ``universal'' player~\cite{DBLP:conf/cav/CoenenFST19}. In this game, the existential and universal players chose the existential and universal traces, respectively, by picking the successor state on paths through the Kripke structure, one position at a time. This strategy can be provided manually or synthesized by a game-solving algorithm. Strategy-based verification is incomplete because the existential player must choose the next state after observing only a finite prefix of the universal traces. The approach can be made complete by adding prophecy variables~\cite{DBLP:conf/csfw/BeutnerF22}. Strategy-based verification was originally developed for finite-state hardware but has been extended to infinite-state software~\cite{DBLP:conf/cav/BeutnerF22}.

\subsubsection*{Bounded Model Checking} In bounded model checking (BMC) for HyperLTL~\cite{10.1007/978-3-030-72016-2_6}, the system is unfolded up to a fixed depth. Pending obligations beyond that depth are either treated pessimistically (to show the satisfaction of a formula) or optimistically (to show the violation of a formula). While BMC for trace properties reduces to SAT solving, BMC for hyperproperties naturally reduces to QBF solving. As usual for bounded methods, BMC for HyperLTL is incomplete.

\subsubsection*{Language inclusion} The complementation operation in the basic algorithm, which is necessary for every quantifier alternation, can be replaced with language inclusion~\cite{tacas-autohyper}. To check a HyperLTL formula $\phi = \forall \pi. \exists \pi'.\, \psi$, we first construct an automaton by intersecting a self-composition of the Kripke structure with the automaton for $\psi$ over the alphabet $(2^\ap)^2$, and then eliminating the second component by existential projection, resulting in an automaton over the alphabet $2^\ap$. The Kripke structure satisfies $\phi$ iff its language is contained in the language of this automaton. Model checking via language inclusion is a complete verification method.

\section{Monitoring}

\begin{figure}
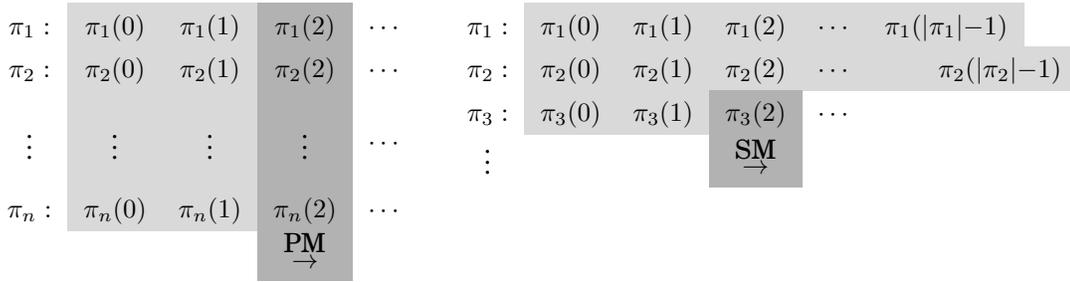

  \centering
  \newcommand\cc[1]{\cellcolor{gray!60}{#1}}
  \newcommand\cl[1]{\cellcolor{gray!30}{#1}} 
  \def\arraystretch{1.5}
\begin{tabular}{cccccccccccccccccccc}
    $\pi_1:$   & \cl{$\pi_1(0)$} & \cl{$\pi_1(1)$} & \cc{$\pi_1(2)$} & $\cdots$ & &   $\pi_1:$   & \cl{$\pi_1(0)$} & \cl{$\pi_1(1)$} & \cl{$\pi_1(2)$} &   \cl{$\cdots$} & \cl{$\pi_1(|\pi_1|{-}1)$} & $\ \ \ $ \\ 
    $\pi_2:$  & \cl{$\pi_2(0)$} & \cl{$\pi_2(1)$} & \cc{$\pi_2(2)$} & $\cdots$ &  &   $\pi_2:$   & \cl{$\pi_2(0)$} & \cl{$\pi_2(1)$} & \cl{$\pi_2(2)$} & \cl{$\cdots$} & \multicolumn{2}{r}{\cl{$\pi_2(|\pi_2|{-}1)$}} & \\ 
   & {\cl{ }} & \cl{ } & \cc{ }  &  &  &   $\pi_3:$   & \cl{$\pi_3(0)$} & \cl{$\pi_3(1)$} & \cc{$\pi_3(2)$} &  $\cdots$ &  &\\
   \multirow{-2}{*}{$\vdots$}  & \multirow{-2}{*}{\cl{$\vdots$}} & \multirow{-2}{*}{\cl{$\vdots$}} & \multirow{-2}{*}{\cc{$\vdots$}} &   \multirow{-2}{*}{$\cdots$}  & &   $\vdots$ &   &        & \cc{$\stackrel{\mbox{SM}}{\rightarrow}$} & & \\
  $\pi_n:$ & \cl{$\pi_n(0)$} & \cl{$\pi_n(1)$} & \cc{$\pi_n(2)$} & $\cdots$ \\
  & &  &  \cc{$\stackrel{\mbox{PM}}{\rightarrow}$} & & \\

\end{tabular}
\caption{Parallel monitoring (PM) vs. sequential monitoring (SM) of hyperproperties. In parallel monitoring, the number of traces is fixed in advance, and the monitor observes one position of all traces at a time; in sequential monitoring, the number of traces is a priori unbounded, and the monitor observes one position of one trace at a time.}
\label{fig:monitoring}
\end{figure}

In monitoring, we determine, at runtime and in an online fashion, whether a system under observation satisfies a given HyperLTL formula. For trace properties, this is typically done by observing growing prefixes of the trace produced by a running system. Monitoring hyperproperties is more difficult because a violation
of a hyperproperty generally involves a set of traces. Two principal ways in which the traces can be presented to the monitor are the parallel and sequential monitoring 
models~\cite{DBLP:journals/fmsd/FinkbeinerHST19} illustrated in Fig.~\ref{fig:monitoring}.
In the \emph{parallel model}, the assumption is that the
number of traces is fixed in advance. All traces then become available
simultaneously, one position at a time from left to right. The
parallel model occurs, for example, in
secure-multi-execution~\cite{conf/sp/DevrieseP10}, where several
system executions are generated by providing different high-security
inputs.  In the \emph{sequential model}, the traces become available
one at a time. In this model, the number of traces is a priori
unbounded.  The sequential model applies, for example, when multiple
sessions of a system are to be monitored one after the other in an
online fashion. In the following, we focus on the sequential model.

\subsection{Finite-trace HyperLTL}

\newcommand{\pathassignfin}{\Pi_\mathit{fin}}
\newcommand{\card}[1]{{|#1|}}

Since the monitoring verdict is based on observing finite traces, we adapt the semantics of HyperLTL to this setting. Let $\pathassignfin \colon \var \rightarrow (2^\ap)^+$ be a partial function mapping trace variables to finite traces. By $t \in \pathassignfin$ we denote that trace $t$ is in the image of $\pathassignfin$.
The satisfaction of a HyperLTL formula $\phi$ over a finite trace assignment $\pathassignfin$ and a set of finite traces $T$, denoted by $(T,\pathassignfin,i) \models \phi$, is defined as follows~\cite{10.1007/978-3-662-54580-5_5,DBLP:conf/tacas/HahnST19}:
\[\def\arraystretch{1.1}
\begin{array}{l@{\hspace{1em}}c@{\hspace{1em}}l}
T,\pathassignfin,i \models a_\pi    & \text{iff} &  a \in \pathassignfin(\pi)(i), \\
T,\pathassignfin,i \models \neg \psi     & \text{iff} &   T,\pathassignfin,i \nmodels \psi, \\
T,\pathassignfin,i \models \psi_1 \wedge \psi_2     & \text{iff} &   T,\pathassignfin,i \models \psi_1 \text{ and } T,\pathassignfin,i \models \psi_2, \\
T,\pathassignfin,i \models \X \psi   & \text{iff} &  \forall t \in \pathassignfin \ldot \card{t}>i+1 \text{ and } T,\pathassignfin,i+1 \models \psi, \\
T,\pathassignfin,i \models \psi_1 \U \psi_2     & \text{iff} &  \exists k \geq i.\ k < \min\limits_{t \in \pathassignfin} \card{t} \wedge T,\pathassignfin,j \models \psi_2\\
& & \land\ \forall i \geq j < k.\ T, \pathassignfin, j \models \psi_1, \\
T,\pathassignfin,i \models \exists \pi \ldot \phi  & \text{iff} & \exists t \in T.\  T,\pathassignfin[\pi \mapsto t],i \models \phi, \\
T,\pathassignfin,i \models \forall \pi \ldot \phi & \text{iff} &  \forall t \in T.\ T,\pathassignfin[\pi \mapsto t],i \models \phi.
\end{array}
\]
Note that the finite traces may differ in length; since the temporal operators advance in all considered traces simultaneously, there is an implicit cut-off at the end of the shortest trace.




%



\subsection{The basic algorithm}

We describe a basic algorithm for sequentially monitoring HyperLTL
formulas of the form $\forall \pi_1, \ldots, \pi_n.\ \psi$, where
$\psi$ is an LTL formula expressing a safety property (for a more
general approach see \cite{DBLP:journals/fmsd/FinkbeinerHST19}). From
$\psi$, we construct a deterministic finite-word automaton
over alphabet $(2^\ap)^n$, which recognizes all sequences
such that the traces $\pi_1, \ldots, \pi_n$, obtained from the
components of the tuples, violate $\psi$. The monitoring algorithm
stores the set $T$ of traces seen so far and a map $S$ that assigns to
each $n$-tuple of traces seen so far a state of the 
automaton.  When a new trace $\pi$ starts, $S$ is initialized with all
$n$-tuples from traces in $T \cup \{\pi\}$ that contain the new trace
$\pi$. When the current trace $\pi$ progresses, we update the state
assigned to each tuple with the successor state of the automaton. If
the automaton reaches a final state, a violation is reported.

\subsection{Practical algorithms}

The currently available algorithms for monitoring hyperproperties can be grouped into two main approaches:

\subsubsection*{Combinatorial approaches} As combinatorial approaches~\cite{DBLP:conf/csfw/AgrawalB16,DBLP:journals/fmsd/FinkbeinerHST19} we understand algorithms that store the traces seen so far and explicitly iterate over the different combinations of the traces. Such approaches can be optimized with efficient data structures for storing sets of traces, such as prefix trees~\cite{DBLP:journals/sttt/FinkbeinerHST20}, and by reducing the number of tuples that need to be considered, for example by recognizing reflexive, symmetric, and transitive specifications. Such optimizations are implemented in the monitoring tool RVHyper~\cite{RVHyper}.

\subsubsection*{Constraint-based approaches}
Constraint-based approaches~\cite{10.1007/978-3-662-54580-5_5,DBLP:conf/tacas/HahnST19} avoid storing the traces explicitly by translating the monitoring task into a constraint system. As new traces arrive, the constraint system is checked for satisfiabilty and rewritten according to the resulting requirements on future traces. 

\section{Synthesis}
\label{sec:synthesis}

\newcommand{\cupdot}{\mathbin{\dot\cup}}
\newcommand{\strat}[2]{(2^{#1})^* \rightarrow 2^{#2}}
\newcommand{\fun}[2]{#1 \rightarrow #2}
\newcommand{\traces}{\mathit{traces}}
\newcommand{\dep}[3]{D^{#3}_{#1 \mapsto #2}}
\newcommand{\arch}{\mathcal{A}}
\newcommand{\penv}{{p_\textit{env}}}
\newcommand{\pminus}{P^{-}}
\newcommand{\tuple}[1]{{\langle #1 \rangle}}


In synthesis, we decide whether there exists an implementation that satisfies a given HyperLTL formula, and, if so, construct such an implementation. We refer to the decision part of synthesis as the \emph{realizability} problem.
We consider HyperLTL formulas where the atomic propositions $\ap = I \cupdot O$ are partitioned into a set $I$ of \emph{inputs} and a set $O$ of \emph{outputs}. We are interested in finding implementations, called \emph{strategies}, which observe the inputs and compute the outputs in such a way that the specification is satisfied.

Formally, a strategy $f \colon \strat{I}{O}$ is a function that maps sequences of input valuations $\pow{I}$ to an output valuation $\pow{O}$.
The behavior of a strategy $f\colon \strat{I}{O}$ is characterized by an infinite tree, called \emph{computation tree}, that branches by the valuations of $I$ and whose nodes $w \in (\pow{I})^*$ are labeled with the strategic choice $f(w)$.
For an infinite word $w = w_0 w_1 w_2 \cdots \in (\pow{I})^\omega$, the corresponding \emph{trace} is defined as $(f(\epsilon) \cup w_0)(f(w_0) \cup w_1)(f(w_0 w_1) \cup w_2)\cdots \in (2^{I \cup O})^\omega$.
The set of traces produced by $f$, written $\traces(f)$, is $\set{w \mid w \in f}$.
We define the satisfaction of a \hyltl formula $\phi$ (over propositions $I \cup O$) on strategy~$f$, written $f \models \phi$, as $\traces(f) \models \phi$.
 A \hyltl formula $\phi$ is \emph{realizable} if there is a strategy $f \colon \strat{I}{O}$ that satisfies $\phi$.

The realizability problem is undecidable. As we will see in the following, already the universal fragment suffices to encode the distributed synthesis problem, which is known to be undecidable~\cite{DBLP:conf/focs/PnueliR90}. The $\exists^*$ and $\exists^*\forall^1$ fragments of HyperLTL are decidable~\cite{DBLP:journals/acta/FinkbeinerHLST20}.
 

\subsection{Distributed synthesis}
 
 Due to the expressiveness of \hyltl, the \hyltl synthesis problem subsumes various classic synthesis problems. For the fragment of HyperLTL with only a single, universal quantifier $\forall \pi.\ \psi$, the \hyltl realizability problem is exactly the LTL realizability problem of $\psi$. With two universal quantifiers, we can express various LTL realizability problems with restrictions on the flow of information, such as synthesis under incomplete information~\cite{conf/ictl/KupfermanV97}, distributed synthesis~\cite{DBLP:conf/focs/PnueliR90}, and fault-tolerant synthesis~\cite{conf/atva/DimitrovaF09}. We illustrate this in the following with the classic distributed synthesis problem (for more details and encodings of other synthesis problems, see~\cite{DBLP:journals/acta/FinkbeinerHLST20}).

 The \emph{distributed synthesis} problem introduces the concept of \emph{architectures} as a constraint on the information flow.
An architecture is a set of processes $P$, with distinct environment process $\penv \in P$, such that the processes produce outputs synchronously, but each process bases its decision only on the history of valuation of inputs that it observes.

\begin{figure}[t]
		\centering
		\begin{tikzpicture}[->,>=stealth',shorten >=1pt,auto,semithick,scale=1,transform shape,scale=0.8]
        \tikzstyle{every state}=[shape=rectangle]
        \tikzstyle{envstate}=[shape=circle,scale=.95]
		
		\node [state,envstate] (e0) {$env$};
		\node [state, below left=1 of e0] (a0) {$p_1$};
		\node [state, below right=1 of e0] (b0) {$p_2$};
		\path[->]
		(e0) edge node [label,above left = 0 and -0.1] {$a$} (a0)
		(e0) edge node [label,above right = 0 and -0.1] {$b$} (b0)
		(a0) edge node [label,above left = -0.15 and 0] {$c$} +(0,-1)
		(b0) edge node [label,above right = -0.15 and 0] {$d$} +(0,-1)
		;
	
		\node [state,envstate, right= 3 of e0] (e) {$env$};
		\node [state, below = of e] (a) {$p_1$};
		\node [state, right=1 of a] (b) {$p_2$};
                \node [state, right=1 of b] (c) {$p_3$};
		\path[->]
		(e) edge node [label,right = 0] {$a$} (a)
		(a) edge node [label,above = 0] {$b$} (b)
		(b) edge node [label,above = 0] {$c$} (c)
		(c) edge node [label,above = 0] {$d$} +(1,0)
		;

                \node [state,envstate, right = 5 of e] (e1) {$env$};
		\node [state, right=1 of e1] (a1) {$p_1$};
		\node [state, right=1 of a1] (b1) {$p_2$};
                \node [state, below=1 of b1] (c1) {$p_3$};
                \node [state, below=1 of a1] (d1) {$p_4$};
                
		\path[->]
		(e1) edge node [label,above = 0] {$a$} (a1)
		(a1) edge node [label,above = 0] {$b$} (b1)
		(b1) edge node [label,right = 0] {$c$} (c1)
		(c1) edge node [label,above = 0] {$d$} (d1)
                (d1) edge node [label,left = 0] {$d$} (a1)
		;

		\end{tikzpicture}

		\caption{Example architectures of distributed systems.} 
                \label{fig:architectures} 
\end{figure}
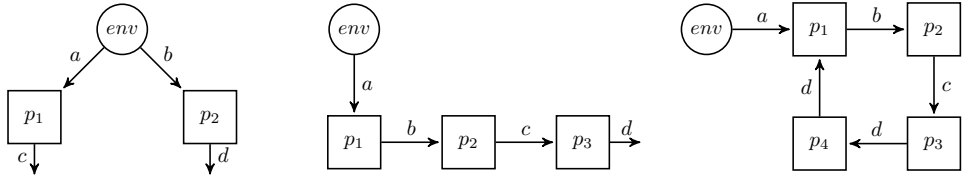

Formally, a distributed architecture $\arch$ is a tuple $\tuple{P, p_{env}, \mathcal{I}, \mathcal{O}}$ where
$P$ is a finite set of processes with distinguished environment process $p_\textit{env} \in P$. The functions $\mathcal{I} \colon \fun{P}{\pow{\ap}}$ and $\mathcal{O} \colon \fun{P}{\pow{\ap}}$ define the inputs and outputs of processes.
While processes may share the same inputs (in case of broadcasting), the outputs of processes must be pairwise disjoint, i.e., for all $p \neq p' \in P$ it holds that $\mathcal{O}(p) \cap \mathcal{O}(p') = \emptyset$.
W.l.o.g.~we assume that $\mathcal{I}(\penv) = \emptyset$.
We denote by $\pminus = P \setminus \set{\penv}$ the set of processes excluding the environment process.
Figure~\ref{fig:architectures} shows several example architectures.


To encode distributed realizability as a \hyltl realizability problem, we first define the notion of \emph{independence} as a hyperproperty: 
\begin{equation*}
\dep{A}{C}{\pi,\pi'} \cceq
  \left(
  \bigvee_{a \in A} (a_\pi \nleftrightarrow a_{\pi'})
  \right)
  \R
  \left(
  \bigwedge_{c \in C} ( c_\pi \leftrightarrow c_{\pi'})
  \right)
\end{equation*}
$\dep{A}{C}{\pi,\pi'}$ requires that the valuations of propositions $C$ on traces $\pi$ and $\pi'$ have to be equal until and including the point in time where there is a difference in the valuation of some proposition in $A$.
Prefacing universal quantification, that is, the formula $\forall \pi \forall \pi' \ldot \dep{A}{C}{\pi,\pi'}$ guarantees that every proposition $c \in C$ solely depends on propositions in $A$.
 Distributed realizability of $\tuple{\varphi,\arch}$ then corresponds to the realizability of the following \hyltl formula over inputs $\mathcal{O}(\penv)$ for $\pminus$ and outputs $\bigcup_{p \in \pminus} \mathcal{O}(p)$: 
  \begin{equation*}
    \forall \pi \forall \pi' \ldot \varphi[\pi] \land \bigwedge_{p \in \pminus} \dep{\mathcal{I}(p)}{\mathcal{O}(p)}{\pi,\pi'}
  \end{equation*}



\subsection{Practical algorithms}

Since the model-checking problem is decidable, the synthesis problem also becomes decidable if the set of potential solutions is finite. Bounded synthesis and controller synthesis ensure this by introducing a bound on the size of the implementation and by fixing the state graph of the implementation, respectively.

\subsubsection*{Bounded synthesis}
In bounded synthesis~\cite{DBLP:journals/sttt/FinkbeinerS13}, the size of the implementation is bounded by a constant, which can be increased incrementally. For universal HyperLTL formulas, this approach has been used to find finite generators of realizing strategies by encoding this search as a satisfiability problem for a decidable constraint system~\cite{DBLP:journals/acta/FinkbeinerHLST20}. In order to detect unrealizable specifications, the approach simultaneously searches for \emph{counterexamples to realizability}. For a universal HyperLTL formula $\phi = \forall \pi_1 \cdots \forall \pi_n \ldot \psi$ over inputs $I$ and outputs $O$, a {counterexample to realizability} is a set of input traces $T \subseteq (2^I)^\omega$ such that for every strategy $f \colon \strat{I}{O}$, the trace set that results from applying $f$ to the input traces in $T$ satisfies $\neg \phi = \exists \pi_1 \cdots \exists \pi_n \ldot \neg\psi$.

\subsubsection*{Controller synthesis}
The controller synthesis problem is a simpler variant of synthesis. In addition to the HyperLTL formula, we are given a \emph{plant model}, which provides the state space of the implementation. The transitions of the plant model are partitioned into controllable and uncontrollable transitions. The goal of the controller synthesis problem is to eliminate a subset of the controllable transitions in such a way that the specification is satisfied. Controller synthesis is decidable for HyperLTL~\cite{DBLP:conf/csfw/BonakdarpourF20}. Controller synthesis has also been combined with synthesis from trace properties. In a first step, the plant is synthesized by finding a (most permissive) strategy for a safety trace property that defines the desired functionality of the system. In a second step, transitions that violate a given hyperproperty are removed from the plant by controller synthesis~\cite{CSF2023}. 

\section{The hierarchy of hyperlogics}
\label{sec:hier}

HyperLTL is the starting point of the hierarchy of \emph{hyperlogics} discussed in the introduction and shown in Fig.~\ref{fig:hier}. This hierarchy was initially studied in~\cite{hierarchy_hyperlogics}; it  adds significant expressiveness to HyperLTL.
Similar to the extension of LTL with trace quantification, the temporal logics HyperQPTL,  HyperCTL$^*$, and HyperQCTL$^*$ are obtained from the standard temporal logics QPTL, CTL$^*$, and QCTL$^*$ by adding quantifiers over traces (or, in the case of CTL$^*$ and QCTL$^*$, paths), so that the formula can refer to multiple traces or paths at the same time. 
The other logics in the hierarchy, FO[$<,E$], S1S[$E$], MPL[$E$], and MSO[$E$] are obtained from the standard first-order and second-order logics FO, S1S, MPL, and MSO by adding the equal-level predicate $E$ (cf.~\cite{Thomas09,Martin}), which indicates that two points happen at the same time.

\subsection{Linear-time hyperlogics}

The logics shown on the left in Fig.~\ref{fig:hier} are linear-time logics.

\subsubsection*{\foe} \foe is a first-order logic for the specification of hyperpropoperties~\cite{Martin}. Kamp’s theorem~\cite{Kamp68} (in the formulation of~\cite{Gabbayetal80}) states that linear-time temporal logic (LTL) is expressively equivalent to monadic first-order logic of order \fo. To express hyperproperties, we add to \fo the equal-level predicate $E$. Given a set of atomic propositions $AP$ and a set $V_1$ of first-order variables, we define the syntax of \foe formulas as follows:
\begin{align*}	
\tau  &\Coloneqq P_a(x) ~|~ x<y ~|~ x=y ~|~ E(x,y) \\
\phi &\Coloneqq \tau ~|~ \neg \phi ~|~ \phi_1 \vee \phi_2  ~|~ \exists x. \phi,
\end{align*}
where $a \in AP$ and $x, y \in V_1$.
%
%
While \fo formulas are interpreted over a trace $t$, we interpret an \foe formula $\phi$ over a set of traces $T$, writing $T \models \phi$ if $T$ satisfies $\phi$. We assign first-order variables with elements from the domain $T \times \nat$. The $<$ relation is defined as the set $\{(t,n_1), (t,n_2) \in (T \times \nat)^2 ~|~ n_1 < n_2 \}$ and the equal-level predicate is defined as $\{ (t_1, n), (t_2, n) \in (T \times \nat)^2\}$.

\foe can express \emph{promptness} requirements such as the existence of a common deadline over all traces by which a certain predicate $a$ must become true on all traces:
\[
\exists x. \forall y.\ E(x,y) \rightarrow P_a(y)
\]
Promptness cannot be expressed in HyperLTL~\cite{BozzelliMP15}. 

\subsubsection*{HyperQPTL} HyperQPTL~\cite{markus} captures the \emph{$\omega$-regular hyperproperties}~\cite{10.1007/978-3-030-53291-8_4}. HyperQPTL extends HyperLTL with quantification over atomic propositions. To easily distinguish quantification over traces $\exists \pi, \forall \pi$ and quantification over propositions $\boldsymbol{\exists} p, {\boldsymbol{\forall}} p$, we use boldface for the latter. The formulas of HyperQPTL are generated by the following grammar:
\begin{align*}\def\arraystretch{1.1}
\phi & {} \cceq {}  \exists \pi.\ \phi ~\mid~ \forall \pi.\ \phi ~\mid~ \psi ~\mid~ \boldsymbol{\exists} p.\ \phi ~\mid~ \boldsymbol{\forall} p.\ \phi \\
\psi & {}  \cceq {}  a_\pi ~\mid~ p ~\mid~ \neg \psi ~\mid~ \psi \wedge \psi ~\mid~ \X \psi ~\mid~ \F \psi 
\end{align*}
where $a,p \in \ap$ and $\pi \in \var$.
	The semantics of HyperQPTL corresponds to the semantics of HyperLTL with additional rules for propositional quantification:
        \[\def\arraystretch{1.1}
\begin{array}{l@{\hspace{1em}}c@{\hspace{1em}}l}
	  T, \pathassign,i  \models \boldsymbol{\exists} q \ldot \phi \quad & \text{iff} &  \exists t \in (2^{\set{q}})^\omega.~ T, \pathassign[\pi_q \mapsto t],i  \models \phi, \\
          T, \pathassign,i  \models \boldsymbol{\forall} q \ldot \phi \quad & \text{iff} & \forall t \in (2^{\set{q}})^\omega.~ T, \pathassign[\pi_q \mapsto t],i  \models \phi, \\
	 T, \pathassign, i \models q  \quad & \text{iff} & \quad q \in \pathassign(\pi_q)(i). 
	 \end{array}
\]

        HyperQPTL can express all properties expressible in \foe, and, additionally, $\omega$-regular hyperproperties like ``on even positions, proposition $a$ has the same value on all traces.'' In the following HyperQPTL formula, even positions are identified by the existentially quantified proposition $\mathit{even}$:
        \[
        \boldsymbol{\exists} \mathit{even}.\, \forall \pi. \forall \pi'.\, \mathit{even} \wedge \G ((\mathit{even} \leftrightarrow \X \neg \mathit{even}) \wedge (\mathit{even} \rightarrow (a_{\pi} \leftrightarrow a_{\pi'})))
        \]
        Another class of properties that can be expressed in HyperQPTL are epistemic properties that express that an agent, who can only observe a subset of the atomic proposition has the \emph{knowledge} that some trace property $\varphi$ is true. This is expressed as the requirement that all traces that agree with the current trace on the observable propositions up to the current point in time satisfy $\varphi$. HyperQPTL subsumes epistemic temporal logic~\cite{markus}. 

\subsubsection*{\seins} \seins is  monadic second-order logic with one successor (S1S) extended with the equal-level predicate.
Let $V_1 = \{x_1, x_2, \ldots\}$ be a set of first-order variables, and $V_2 = \{X_1, X_2, \ldots\}$ a set of second-order variables.
The formulas $\phi$ of \seins are generated by the following grammar:
\begin{align*}
\tau &\Coloneqq x ~\mid~ \mathit{min}(x) ~\mid~ \mathit{Succ}(\tau)\\
\phi &\Coloneqq \tau \in X ~\mid~ \tau = \tau ~\mid~ E(\tau,\tau) ~\mid~ \neg \phi ~\mid~ \phi \vee \phi ~\mid~ \exists x. \phi ~\mid~ \exists X. \phi,
\end{align*}
where $x \in V_1$ is a first-order variable, $\mathit{Succ}$ denotes the successor relation, and $\mathit{min}(x)$ indicates the minimal element of the trace identified by $x$.
Furthermore, $E(\tau, \tau)$ is the equal-level predicate and $X \in V_2 \cup \{X_a ~|~ a \in AP \}$.
%
We interpret \seins formulas over a set of traces $T$. 
As for \foe, the domain of the first-order variables is $T \times \nat$.

\seins adds significant expressiveness to HyperQPTL. For example, it is possible to encode the existence of a terminating computation of a Turing machine~\cite{hierarchy_hyperlogics}. While the model-checking problem of HyperQPTL is still decidable (with a similar algorithm as for HyperLTL), the model-checking problem for \seins is thus undecidable.

\subsection{Branching-time hyperlogics}

The logics shown on the right in Fig.~\ref{fig:hier} are branching-time logics. While the linear-time logics are interpreted over sets of traces, we interpret the branching-time logics over infinite trees. A Kripke structure satisfies a branching-time formula iff its unrolling into an infinite tree satisfies the formula.

\subsubsection*{HyperCTL$^*$} Extending HyperLTL to branching time leads to the temporal logic HyperCTL$^*$. HyperCTL$^*$~\cite{DBLP:conf/post/ClarksonFKMRS14} has the same syntax as HyperLTL, except that the quantifiers refer to paths rather than traces, and that path quantifiers may occur in the scope of temporal modalities. Let $\pi \in \mathcal{V}$ be a \emph{path variable} from an infinite supply of path variables $\mathcal{V}$ and let $\exists \pi.~\phi$ be the explicit existential \emph{path quantification}. 
HyperCTL$^*$ formulas are generated by the following grammar:
\[
\phi \Coloneqq a_\pi \mid \neg \phi \mid \phi \vee \phi \mid \X \phi \mid \phi \U \phi \mid \exists \pi.~ \phi.
\]
The semantics of a HyperCTL$^*$ formula is defined with respect to an infinite tree-shaped Kripke structure $\mathcal{T}$ and a \emph{path assignment} $\Pi:\mathcal{V} \rightarrow \mathit{Paths}(\mathcal{T})$, which is a partial mapping from path variables to paths in the tree. 
The satisfaction relation $\models$ is given as follows:
\[\def\arraystretch{1.1}
\begin{array}{l@{\hspace{1em}}c@{\hspace{1em}}l}
\mathcal{T}, \Pi, i \models a_\pi & \text{iff} & a \in L(\Pi(\pi)(i)), \\
\mathcal{T}, \Pi, i \models \neg \phi & \text{iff} & \Pi, i \nmodels_\mathcal{T} \phi, \\
\mathcal{T}, \Pi, i \models \phi_1 \wedge \phi_2 & \text{iff} & \mathcal{T}, \Pi, i \models \phi_1 \text{ and } \mathcal{T}, \Pi, i \models \phi_2, \\
\mathcal{T}, \Pi, i \models \X \phi & \text{iff} & \mathcal{T}, \Pi, i + 1 \models \phi,\\
\mathcal{T}, \Pi, i \models \phi_1 \U \phi_2 & \text{iff} & \exists j \geq i.~ \mathcal{T}, \Pi, j \models \phi_2,\\
&& \land\ \forall i \leq k < j. ~\mathcal{T}, \Pi, k \models_\mathcal{T} \phi_1,\\
\mathcal{T}, \Pi, i \models \exists \pi. \phi & \text{iff} & \exists p \in \mathit{Paths}(\mathcal{T}).~ p[0, i] = \varepsilon[0, i], \\
&& \land\ \mathcal{T}, \Pi[\pi \mapsto p, \varepsilon \mapsto p], i \models \phi,
\end{array}
\]
where we use $\varepsilon$ to denote the last path that was added to the path assignment $\Pi$. We say that a tree $\mathcal{T}$ satisfies a HyperCTL$^*$ formula $\phi$, written as $\mathcal{T} \models \phi$, if $\mathcal{T}, \emptyset,0 \models \phi$, where $\emptyset$ denotes the empty path assignment.
HyperCTL$^*$ can express the flow of information that appears in different branches of the computation tree, as illustrated in the following example (taken from \cite{journals/eatcs/Finkbeiner17}):

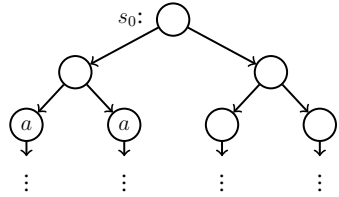
\begin{center}
\begin{tikzpicture}[level distance=7mm,line width=0.75pt,
    level 1/.style={sibling distance=26mm,
      edge from parent/.style={draw,->,solid,line width=0.75pt}},
    level 2/.style={sibling distance=13mm,
      edge from parent/.append style={draw,->,solid,line width=0.75pt}},
    level 3/.style={sibling distance=10mm,
      edge from parent/.append style={draw,->,solid,line width=0.75pt}}]
		\tikzstyle{every node}=[font=\large]
    \node [scale=0.7] (a) {$s_0$:\qquad\quad~~~~}  
				node [circle,draw,scale=0.7] (a) {\phantom{$a$}} 
        child {node [circle,draw,scale=0.7] (a1) {\phantom{$a$}}
          child {node [circle,draw,scale=0.7] (a11) {$a$}
          child {node [scale=0.7] (a1111) {\vdots}}}
          child {node [circle,draw,scale=0.7] (a21) {$a$}
            child {node [scale=0.7] (a1111) {\vdots}}}
      }
        child {node [circle,draw,scale=0.7] (a1) {\phantom{$a$}}
          child {node [circle,draw,scale=0.7] (a11) {\phantom{$a$}}
            child {node [scale=0.7] (a1111) {\vdots}}}
          child {node [circle,draw,scale=0.7] (a21) {\phantom{$a$}}
            child {node [scale=0.7] (a1111) {\vdots}}}
	};
\end{tikzpicture}
\end{center}

An observer who sees $a$ can infer which branch was taken in the first nondeterministic choice, but not which branch was taken in the second nondeterministic choice. This is expressed by the HyperCTL$^*$ formula \[\forall \pi.\, \X \forall \pi' .\, \X (a_\pi \equ a_{\pi'}).\]

\subsubsection*{\mple}
Monadic path logic equipped with the equal-level predicate (\mple)
is the extension of \foe with second-order quantification, where the second-order quantification is restricted to full paths in the tree.

Let $V_1 = \{x_1, x_2, \ldots\}$ be a set of first-order variables, and $V_2 = \{X_1, X_2, \ldots\}$ a set of second-order variables.
The formulas $\phi$ of \mple are generated by the following grammar: 
\begin{align*}	
  \phi &\Coloneqq \psi ~|~ \neg \phi ~|~ \phi_1 \vee \phi_2  ~|~ \exists x. \phi ~|~ \exists X. \phi\\
\psi  &\Coloneqq P_a(x) ~|~ x<y ~|~ x=y ~|~ x\in X ~|~ E(x,y), 
\end{align*}
where $a \in AP$, $x, y \in V_1$, and $X \in V_2$.
In the semantics of \mple, first-order variables are assigned to nodes in the tree and second-order variables to sets of nodes whose elements constitute exactly one path of the tree. $P_a$ evaluates to the set of nodes whose label contains $a$.
$x < y$ indicates that $x$ is an ancestor of $y$. 
Atomic formulas $x \in X$ and $x = y$ are interpreted as set membership and equality on nodes, respectively.
The equal-level predicate $E(x,y)$ denotes that two nodes $x$ and $y$ are on the same level, i.e., have the same number of ancestors.
\mple is strictly more expressive than HyperCTL$^*$. Similarily to HyperQPTL, \mple can express properties about knowledge~\cite{hierarchy_hyperlogics}.

\subsubsection*{HyperQCTL$^*$ and \msoe} HyperQCTL$^*$~\cite{hierarchy_hyperlogics} extends HyperCTL$^*$ with quantification over atomic propositions.
The formulas of Hyper\-QCTL$^*$ are generated by the following grammar:
\[
\begin{array}{llllllllllllllllll}
\phi ~&~::= ~&~ a_{\pi} ~&~ \vert ~&~ \neg\phi ~&~ \vert ~&~ 
\phi\vee\phi ~&~ \vert ~&~ \X\phi ~&~ \vert ~&~ \phi \U\phi ~&~ \vert ~&~ \exists \pi.\; \phi ~&~ \vert ~&~ \boldsymbol{\exists} p.\; \phi
\end{array}
\]
where $a,p \in \ap$ and $\pi \in \var$.
The semantics of HyperQCTL$^*$ corresponds to the semantics of HyperCTL$^*$ with an additional rule for propositional quantification: $\boldsymbol{\exists} p. \phi$ is satisfied iff $\phi$ is satisfied after the labeling of the tree has been modified by (re-)assigning the valuation of $p$ to the nodes of the tree.
%
%
%
HyperQCTL$^*$ has the same expressiveness as second-order monadic logic equipped with the equal-level predicate (MSO[$E$]), i.e., the extension of \foe with second-order quantification~\cite{hierarchy_hyperlogics}.
While the model-checking problem for \mple is still decidable~\cite{DBLP:conf/vmcai/Finkbeiner21}, the model-checking problems for HyperQCTL$^*$ and \msoe are undecidable~\cite{hierarchy_hyperlogics}.

\subsection{Further extensions}

In addition to the logics discussed so far, HyperLTL has been extended in numerous  further directions. In the following, we briefly discuss three major areas of ongoing research.

\subsubsection*{Probabilistic hyperproperties}

To reason about probabilistic hyperproperties of Markov decision processes, HyperLTL has been extended with probabilistic operators and strategy quantifiers, resulting in the temporal logic \emph{PHL}~\cite{DimitrovaFT20};
similar extensions are the logics \emph{HyperPCTL}~\cite{DBLP:conf/qest/AbrahamB18,AbrahamBBD20} and \emph{HyperPCTL$^*$}~\cite{DBLP:conf/csfw/0044NBP21}. 

\subsubsection*{Infinite-state hyperproperties}

\emph{HyperTSL}~\cite{CSF2023} is a temporal logic for the specification of hyperproperties of infinite-state software. It is based on temporal stream logic (TSL)~\cite{DBLP:conf/cav/Finkbeiner0PS19}, which extends linear temporal logic (LTL) with the concept of cells and uninterpreted functions and predicates. These mechanisms separate the temporal control flow from the concrete data in the system, which enables reasoning about infinite-state systems. 
\emph{First-order
HyperLTL} extends HyperLTL with first-order quantifiers over uninterpreted sorts~\cite{DBLP:conf/ccs/Finkbeiner0SZ17}. 
For cyber-physical systems, HyperLTL has also been extended with \emph{real-valued signals}~\cite{10.1145/3127041.3127058,DBLP:journals/tecs/0044ZBP19}. 

\subsubsection*{Asynchronous hyperproperties}

While HyperLTL has a strictly
synchronous semantics, and a general logic for asynchronous
hyperproperties has not been proposed yet, there are already two
extensions of HyperLTL that go in this direction: the extension with
\emph{trajectory quantifiers}~\cite{DBLP:conf/cav/BaumeisterCBFS21}
considers multiple schedulings between the traces; \emph{Stuttering
HyperLTL}~\cite{DBLP:conf/lics/BozzelliPS21} eliminates stuttering steps
from the traces. A further big step towards asynchronous
hyperproperties is the \emph{temporal fixpoint calculus}
$H_\mu$~\cite{10.1145/3434319}, which integrates the quantifiers from
HyperLTL into the modal $\mu$-calculus.
Many asynchronous hyperproperties can also be encoded in \emph{HyperATL$^*$}~\cite{DBLP:conf/concur/BeutnerF21}, an extension of HyperLTL to games with multiple coalitions of players. 
Asynchronous hyperproperties are increasingly supported by practical verification approaches such as bounded model checking~\cite{tacas-borzoo}.

\section{Concluding Remarks}

HyperLTL and its extensions are a powerful unifying framework for reasoning about hyperproperties. For a wide range of algorithmic problems, including satisfiability, model checking, monitoring, and synthesis, the logics provide a common specification language and semantic foundation. It is encouraging that the model-checking problem remains decidable for a large portion of the hierarchy of hyperlogics, allowing for the development of automated verification techniques that cover broad classes of hyperproperties in a uniform manner.

Despite the significant progress over the past decade, much work remains to be done. Many of the currently available algorithms focus on finite-state hardware; reasoning about hyperproperties of infinite-state software remains a challenge. It is also clear that the expressiveness of logics like HyperLTL is dwarfed by the general power of hyperproperties. Many concepts that relate to concerns like privacy~\cite{KalloniatisKG08}, explainability~\cite{596837f7d6814b3b9e2ca0a5d9928aa3}, or fairness~\cite{MehrabiMSLG21} are arguably covered by Clarkson and Schneider's general definition of hyperproperties as properties of sets of traces; yet a specification language that would provide a unifying theory for all these notions remains out of reach for now.

%
%
%
%
%
%
%
%
\begin{acks}
  This work was partially supported by the European Research Council (ERC) grant HYPER (No. 101055412).
\end{acks}

\bibliographystyle{ACM-Reference-Format-Journals}

\bibliography{bibliography,biblio,hyper} 

\end{document}